# Adaptive Cache Management for Complex Storage Systems Using CNN-LSTM-Based Spatiotemporal Prediction

Xiaoye Wang
Western University
London, Canada

Xuan Li
Columbia University
New York, USA

Linji Wang
George Mason University
Fairfax, USA

Tingyi Ruan
Northeastern University
Boston, USA

Pochun Li*
Northeastern University
Boston, USA

*Abstract*—This paper proposes an intelligent cache management strategy based on CNN-LSTM to improve the performance and cache hit rate of storage systems. Through comparative experiments with traditional algorithms (such as LRU and LFU) and other deep learning models (such as RNN, GRU-RNN and LSTM), the results show that the CNN-LSTM model has significant advantages in cache demand prediction. The MSE and MAE values of this model are significantly reduced, proving its effectiveness under complex data access patterns. This study not only verifies the potential of deep learning technology in storage system optimization, but also provides direction and reference for further optimizing and improving cache management strategies. This intelligent cache management strategy performs well in complex storage environments. By combining the spatial feature extraction capabilities of convolutional neural networks and the time series modeling capabilities of long short-term memory networks, the CNN-LSTM model can more accurately predict cache needs, thereby Dynamically optimize cache allocation to improve system response speed and resource utilization. This research provides theoretical support and practical reference for cache optimization under large-scale data access modes, and is of great significance to improving the performance of future storage systems.

## I. Introduction

With the development of big data technology and cloud computing, the scale and complexity of storage systems are increasing, and the storage, management and access requirements of data are becoming more and more complex. In order to improve the performance and resource utilization efficiency of storage systems, intelligent cache management strategies have become a key research direction. Traditional cache management strategies mostly rely on fixed rules and algorithms, such as LRU (least recently used) and LFU (least frequently used). Although these strategies work well in simple environments, they have limitations when faced with complex and changing data access patterns [1]. Therefore, exploring more intelligent cache management strategies to adapt to large-scale data environments and complex storage access patterns has become an important and urgent topic .

With the advancement of artificial intelligence and deep learning technologies, intelligent cache management strategies have a new development direction [2]. Compared with traditional rules and heuristic algorithms, deep learning can learn hidden features from complex data patterns and make predictions based on historical access records[3-5]. The combination of CNN-LSTM model is a typical application that can fully utilize the advantages of convolutional neural network (CNN) in feature extraction and the powerful ability [6] of long short-term memory network (LSTM) in time series data prediction. CNN can extract spatial features from historical data access patterns, such as data access frequency and access preference distribution, while LSTM can learn the temporal characteristics of data access, such as data that may be accessed again at a certain moment [7]. Therefore, the CNN-LSTM model can accurately predict future cache requirements and provide strong support for the cache management strategy of the storage system.

In current storage systems, cache hit rate directly affects the overall performance of the system [8]. A higher cache hit rate can reduce access latency and I/O operation load, thereby improving the system's response speed and throughput. This has a significant impact on many I/O-intensive fields, where fast data retrieval and processing are essential for optimal performance. For instance, in UI interaction systems [9], a higher cache hit rate enables smoother user experiences by reducing delays in displaying content. In medical imaging [10-12], quick access to cached data can expedite image processing and analysis, which is critical for timely diagnoses and treatment decisions. Similarly, in various online services[13-15], improved cache performance reduces server load and enhances responsiveness, enabling these platforms to handle higher traffic volumes while maintaining a seamless user experience. However, due to the wide variety of data and diverse access patterns in storage systems, optimizing cache hit rate often faces huge challenges. Traditional cache management strategies are usually based on past experience and rules. Although this method is simple, it is difficult to achieve accurate prediction. Through artificial intelligence technology, we can learn from historical data and generate more accurate and personalized cache strategies. In particular, through the CNN-LSTM model, we can learn based on the spatiotemporal characteristics of data without relying on fixed rules [16], dynamically adjust the cache strategy, and thus improve the cache hit rate.

Another advantage of using the CNN-LSTM model for intelligent cache management is its scalability and adaptability [17]. The workload and data access patterns of storage systems are often dynamically changing, and traditional static rules often perform poorly in the face of changes. The CNN-LSTM model can continuously update and adjust its internal parameters over time to adapt to new data access patterns. This adaptive capability enables storage systems to maintain efficient and stable operation when facing complex and dynamic workloads. At the same time, the deep learning-based model can continuously improve prediction accuracy as the amount of data increases, providing the possibility of further optimizing cache management strategies.

## II. METHOD

The paper presents a cache demand prediction method based on a CNN-LSTM framework to enhance system cache hit rates. The core idea is to employ CNN to extract features from historical data, followed by LSTM to model and predict these features over time, ultimately generating forecasts for future cache demand. These predictions enable dynamic adjustments to the cache management strategy. The CNN component extracts spatial features from data access patterns [18], such as access frequency and preference distribution, while the LSTM component captures temporal characteristics, identifying trends in cache demand. By deeply analyzing historical data, this method helps maintain high cache hit rates under varying workloads, reducing access latency and optimizing system response speed and throughput.

Unlike traditional cache management strategies, the CNN-LSTM model does not rely solely on fixed rules, but automatically learns features from data through deep learning, so that the system can adapt to different access patterns more accurately. The prediction method leveraging spatiotemporal features significantly enhances the accuracy and intelligence of cache management. This approach is substantiated by the work of Yan et al. [19], whose findings offer a foundational framework for applying spatiotemporal modeling to effectively manage complex cache behaviors. Their research facilitates more precise and proactive cache management strategies, particularly valuable in diverse, data-intensive environments. The overall architecture of the model is shown in Figure 1.

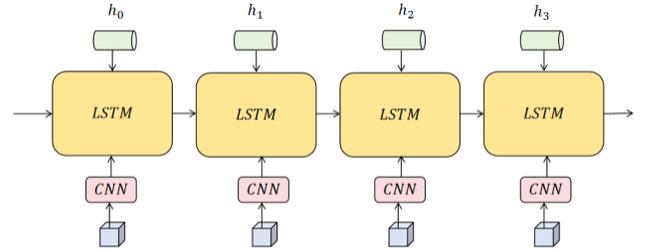

Figure 1 Overall architecture of the model

First, we collect historical data from the storage system, including features such as data access frequency, data block size, access latency, and cache hit rate in each time period. These features form a multi-dimensional feature matrix of time series, denoted as $X \in R^{T \times N}$, where $T$ represents the number of time steps and $N$ represents the dimension of the feature. We input this multi-dimensional feature matrix into the model to learn the spatiotemporal characteristics of the storage system.

In the feature extraction stage [20], we use convolutional neural network (CNN) to extract the spatial features of the data. Assuming the input feature matrix is $X$, after a convolution layer, the output feature matrix $X'$ is:

$$X' = f(W * X + b)$$

Among them, $W$ is the convolution kernel, $b$ is the bias term, $*$ represents the convolution operation, and $f(\cdot)$ is the activation function. Through multi-layer convolution operations [21-22], we can get a more refined feature representation. The purpose of this stage is to extract representative features from multi-dimensional features to reduce the computational complexity of the model while retaining key information.

Next, we input the feature $X'$ output by the convolutional layer into the LSTM network. The LSTM network can model time series data and effectively capture long-term dependencies. The basic unit of LSTM can be expressed as:

$$\begin{cases} i_t = \sigma(W_i \cdot [h_{t-1}, X_t'] + b_i) \\ f_t = \sigma(W_f \cdot [h_{t-1}, X_t'] + b_f) \\ o_t = \sigma(W_o \cdot [h_{t-1}, X_t'] + b_o) \\ c_t = f_t \otimes c_{t-1} + i_t \otimes \tanh(W_c \cdot [h_{t-1}, X_t'] + b_c) \\ h_t = o_t \otimes \tanh(c_t) \end{cases}$$

Among them, $i_t$, $f_t$ and $o_t$ are the input gate, forget gate and output gate respectively, $c_t$ is the state of the memory unit,

and $h_t$ is the hidden state of the current time step. $\sigma$ represents the sigmoid function and $\otimes$ represents the element-level multiplication operation. The structure of LSTM allows the network to capture the long-term pattern of data access over a long time series and predict future cache requirements.

Through the combination of CNN and LSTM, we get a series of hidden states, which contain all the information of past features in the time series. We map these hidden states to a scalar output through a fully connected layer, indicating the cache requirement at a certain time step in the future:

$$y_t = W_y \cdot h_t + b_y$$

Where $W_y$ and $b_y$ are the weights and biases of the fully connected layer. The output $y_t$ represents the predicted value of cache demand at time $t$.

In order to optimize the parameters of the model, we use Mean Squared Error (MSE) as the loss function, which is defined as:

$$L = \frac{1}{N} \sum_{t-1}^{T} (y_t - y'_t)^2$$

Among them, $y'_t$ is the actual cache demand value. By minimizing this loss function, we can adjust the parameters of the model to make the predicted value as close to the actual value as possible.

After the model training is completed, we use the prediction results to optimize the storage system's cache strategies. Specifically, we determine which data blocks will be frequently accessed based on the predicted future cache demand $y_t$, so that these data blocks can be loaded into the cache in advance to improve the cache hit rate. At the same time, for the predicted low-access frequency data blocks, they can be migrated to the secondary storage to save high-performance storage space. In addition, by adjusting the size and priority of the cache, the model can dynamically adapt to system changes, thereby optimizing overall performance.

The advantage of the entire method is that it combines the spatial feature extraction capabilities of CNN and the time series modeling capabilities of LSTM to make the prediction of cache requirements more accurate. The spatial characteristics of data access, such as access frequency and preference distribution, are extracted through CNN, and LSTM performs time series analysis on these characteristics to capture the trend of cache demand changing over time. This combination enables the model to gain a deep understanding of the spatiotemporal characteristics of caching requirements, thereby providing more accurate predictions in changing data access patterns.

In addition, the method is dynamically adaptable and can continuously update model parameters according to different data access patterns and storage system states, allowing the system to adapt to changing workloads and access requirements. Through this dynamic adjustment, the system can maintain efficient and stable performance in complex storage environments, improve cache hit rates, and reduce access delays. This adaptive feature allows this method to still achieve high resource utilization and response efficiency when processing large-scale data, thereby significantly optimizing the overall performance of the storage system.

III. EXPERIMENT

A. Datasets

The dataset used in this article is the Traces dataset from Microsoft Research, which is a real storage system access log dataset. This dataset records the user's access to the storage system on multiple servers, including access timestamps, file paths, operation types (such as read, write), file sizes and other detailed information. This dataset covers a large number of operations and has a long recording time span, which is very suitable for studying the cache management strategy of the storage system.

By analyzing access logs, we can uncover the access patterns and characteristics of the storage system, providing valuable data for model training and optimization. In cache management, this data enables a deeper understanding of access frequency, timing characteristics, and load distribution across different data types, thereby offering a reliable basis for cache demand prediction. Leveraging this information, the model can accurately identify trends in high-frequency data access and cache demand, allowing the system to adapt more effectively to varying workloads. This optimization of cache allocation strategies enhances cache hit rates, reduces access latency, and ultimately supports the efficient operation of the system.

B. Experiments

In addition to the CNN-LSTM model proposed in this paper, we also selected the following five models for comparative experiments: the traditional LRU (least recently used) algorithm, which caches the most recently used data blocks through simple rules; the LFU (least frequently used) algorithm, which determines the cache strategy based on the access frequency; the RNN-based cache prediction model [23], which can capture short-term features in time series; the GRU-CNN combination model, which uses CNN to extract spatial features and combines GRU to capture time series information; and the simple LSTM model, which has certain advantages in time series modeling.

The selection of these models covers cache management strategies from traditional algorithms to multiple deep learning methods. By comparing their performance in cache hit rate and access latency, the superiority of the CNN-LSTM model can be more comprehensively verified. The experimental results are shown in Table 1.

Table 1 Experiment result

| Model | MSE | MAE |
|---|---|---|
| LRU | 0.951 | 0.867 |
| LFU | 0.873 | 0.725 |
| RNN | 0.623 | 0.546 |
| GRU-RNN | 0.521 | 0.465 |
| LSTM | 0.375 | 0.321 |
| CNN-LSTM(Ours) | 0.244 | 0.127 |

It can be seen from the experimental results that there are significant differences in the performance of different models in the cache demand prediction task. We use MSE (Mean Square Error) and MAE (Mean Absolute Error) as evaluation indicators to measure the prediction accuracy of each model. The overall trend shows that from the simplest LRU and LFU algorithms to more complex deep learning models, the prediction performance gradually improves, and the CNN-LSTM model proposed in this article performs the best.

First, analyze two traditional cache management algorithms, LRU and LFU. Their MSEs are 0.951 and 0.873 respectively, and their MAEs are 0.867 and 0.725 respectively. These algorithms rely on the most recent or most frequent usage of cache blocks to manage the cache. Although they are simple and computationally cost-effective, they do not consider the spatiotemporal characteristics of data access, so their caching strategies are less accurate and therefore have larger errors in complex storage system environments.

Next is the RNN model with an MSE of 0.623 and a MAE of 0.546. Compared with traditional algorithms, RNN models can capture short-term dependency characteristics in data time series. The introduction of RNN shows that even a simple neural network structure can significantly improve the performance of cache demand prediction. However, the disadvantage of RNN is its weak ability to handle long-term dependencies. When the data access pattern is more complex or has large long-term changes, the prediction accuracy of RNN is still limited, so its error value is still high in deep learning models.

The GRU-RNN model further improves the performance, with MSE and MAE of 0.521 and 0.465 respectively. GRU (gated recurrent unit) is an improved version of RNN, which introduces a gating mechanism to solve the vanishing gradient problem of RNN in long-term dependency learning. This enables GRU-RNN to better capture changes in long-term access patterns, improve prediction accuracy while reducing computational costs. Therefore, compared with RNN, the error value of GRU-RNN is significantly reduced, proving the effectiveness of the gating mechanism in cache demand prediction.

The LSTM model performs better than GRU-RNN, with an MSE of 0.375 and a MAE of 0.321. This is because LSTM (Long Short-Term Memory Network) adds forget gates, input gates, and output gates to the structure, which can better control the flow of information and thus perform well on long-term sequence data. Compared with GRU, the model complexity of LSTM is slightly increased, but its ability to capture long-term dependencies is stronger, so it performs better in cache demand prediction.

Finally, the CNN-LSTM model proposed in this article achieved the best experimental results, with an MSE of 0.244 and a MAE of 0.127. This is because we combine the advantages of CNN and LSTM. CNN is responsible for extracting the spatial characteristics of data, such as the pattern and frequency distribution of data access; while LSTM processes time series information and effectively models the changing trend of data over time. By combining these two network structures, the model can make full use of the spatiotemporal characteristics of the data to make cache demand prediction more accurate. Especially when the data access pattern has obvious spatiotemporal characteristics, the advantages of the CNN-LSTM model are fully utilized and the error value is significantly reduced.

The experimental results further verified the potential of the deep learning model in storage system optimization. Since traditional algorithms rely only on simple rules, their performance in complex environments is limited. As the complexity and intelligence of the model increases, the model can better adapt to changing data access patterns. In particular, deep learning models, such as LSTM and CNN-LSTM, are not only able to handle complex temporal dependencies, but also capture hidden patterns in data through feature extraction modules (such as CNN), significantly improving prediction accuracy.

Finally, we also show images of training loss reduction, as shown in Figure 2.

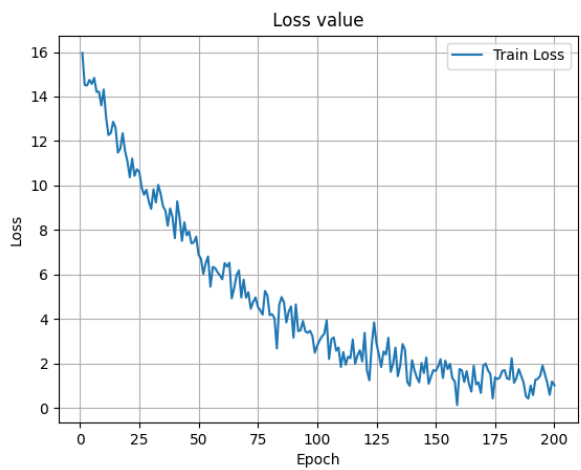

Figure 2 Training loss drop graph

## IV. Conclusion

This article summarizes the research results on the optimization of cache management in storage systems based on the CNN-LSTM model. Through comparative experiments with other models, we verified the superiority of this model in cache demand prediction. Traditional LRU and LFU algorithms have limited performance when facing complex data access patterns, while deep learning models such as RNN, GRU-RNN and LSTM significantly improve prediction accuracy. However, the CNN-LSTM model proposed in this article combines the feature extraction capabilities of convolutional neural networks and the time series modeling capabilities of long short-term memory networks, which can better capture the spatiotemporal characteristics of data access. Therefore, the experimental results show that the CNN-LSTM model performs best in cache demand prediction, effectively improving cache hit rate and system performance.

The significance of this study is to demonstrate the potential of deep learning technology in storage system optimization, especially the application of intelligent cache management in complex and dynamically changing storage environments. While the CNN-LSTM model effectively predicts cache demands, it has limitations. Its performance can depend heavily on hardware, making it less suitable for low-resource environments. Future research can continue to explore multi-modal data fusion and more complex deep learning architectures to further improve the prediction accuracy and adaptability of the model. This not only helps improve the overall performance of the storage system, but also provides important technical support and practical reference for large data centers and cloud service providers in data storage management.